\documentstyle[sprocl]{article}
\def\Journal#1#2#3#4{{#1} {\bf #2}, #3 (#4)}

\def\PRL{\em Phys. Rev. Lett.}

\def\be{\begin{equation}}
\def\ee{\end{equation}}
\def\bea{\begin{eqnarray}}
\def\eea{\end{eqnarray}}

\def\dzeroh{\hat\partial_0}
\def\Boxh{\hat{\mbox{\kern-.0em\lower.3ex\hbox{$\Box$}}}}
\def\bK{{\bf K}}
\def\bg{{\bf g}}

\begin{document}
\title{HYPERBOLIC FORMULATION OF GENERAL RELATIVITY}
\author{ ANDREW ABRAHAMS\footnote{Permanent address: 
NCSA, Beckman Institute, University of Illinois, Champaign IL 61820 USA}, 
ARLEN ANDERSON, 
YVONNE CHOQUET-BRUHAT\footnote{Permanent address: Gravitation et Cosmologie
Relativiste, t.22-12, Un. Paris VI, Paris 75252 France.}, and JAMES W. YORK,
JR.
}
\address{Dept. Physics and Astronomy\\ Univ. North Carolina\\
Chapel Hill, NC 27599-3255 USA}

\maketitle\abstracts{
Two geometrical well-posed hyperbolic formulations of general 
relativity are described. One admits any time-slicing which 
preserves a generalized harmonic condition.  The other admits
arbitrary time-slicings.  Both systems have only the 
physical characteristic speeds of zero and the speed of light.
}

\vspace{-2.8in}
\hfill IFP-UNC-520

\hfill TAR-UNC-055

\hfill gr-qc/9703010
\vspace{2.1in}

Einstein's theory, viewed mathematically as a system of second-order partial 
differential equations for the metric, is not a hyperbolic system 
without modification and is not manifestly well-posed, though
physical information propagates at the speed of light.
A well-posed hyperbolic system admits unique solutions which
depend continuously on the initial data and seems to 
be required for robust, stable numerical integration. 
The well-known traditional approach achieves hyperbolicity
through special coordinate choices.  The formulations\cite{aacby} described 
here permit much greater coordinate gauge freedom.  Because these
exact nonlinear theories incorporate the constraints, they are 
natural starting points for developing gauge-invariant perturbation 
theory. 

Consider a globally hyperbolic manifold of topology $\Sigma\times R$ 
with the metric 
\begin{equation}
ds^2 = -N^2 dt^2 +g_{ij} (dx^i +\beta^i dt) (dx^j +\beta^j dt),
\end{equation}
where $N$ is the lapse, $\beta^i$ is the shift, and $g_{ij}$ is the
spatial metric.  The derivative, 
$N^{-1}\hat\partial_0=N^{-1}(\partial/\partial t -{\cal L}_{\beta})$,
where ${\cal L}_{\beta}$ is the Lie derivative in a time slice $\Sigma$ along
the shift vector, is the derivative with respect to proper time
along the normal to $\Sigma$. This implies that $\dzeroh$ is the natural 
time derivative for evolution.
The extrinsic curvature $K_{ij}$ of $\Sigma$ is defined as
\begin{equation}
\label{Kij}
\hat\partial_0 g_{ij}= -2 N K_{ij}.
\end{equation}

Together with (\ref{Kij}), the dynamical part of Einstein's theory 
(in vacuum) can be expressed in 3+1 language as
\be
\label{Rij}
R_{ij}=-N^{-1} \dzeroh K_{ij} +{\jmath}_{ij}+ \bar R_{ij}=0,
\ee
where $\bar R_{ij}$ is the spatial Ricci curvature and ${\jmath}_{ij}$
consists of terms at most zeroth order in derivatives of $\bK$,
first order in derivatives of $\bg$, and second order in derivatives of
$N$.  The spatial Ricci curvature is second order in derivatives of $\bg$
in such a way as to spoil the hyperbolicity of the combined equations
(\ref{Kij}) and (\ref{Rij}). 
To achieve hyperbolicity for the 3+1 equations, we proceed as follows.

By taking a time derivative of (\ref{Rij}) and subtracting 
appropriate spatial covariant derivatives of the momentum constraints,
we obtain an equation with a wave operator acting on the extrinsic
curvature.  In vacuum, one finds
\be
\label{boxK}
\dzeroh R_{ij} -\bar \nabla_i R_{0j} -\bar\nabla_j R_{0i}=
N\Boxh K_{ij}  + J_{ij} + S_{ij}=0,
\ee
where $\Boxh=-N^{-1}\dzeroh N^{-1}\dzeroh + \bar\nabla^k \bar\nabla_k$,
$J_{ij}$ consists of terms at most first order in derivatives of $\bK$,
second order in derivatives of $\bg$, and second order in derivatives of
$N$, and
\be
\label{Sij}
S_{ij}=-N^{-1} \bar\nabla_i \bar\nabla_j (\dzeroh N + N^2 H)
\ee
($H=K^k\mathstrut_k$).  The term $S_{ij}$ is second order in derivatives
of $\bK$ and  would spoil hyperbolicity of the wave operator
$\Boxh$ acting on $\bK$.  Hyperbolicity is achieved by imposing ``generalized
harmonic slicing''
\be
\label{dzeroN}
\dzeroh N + N^2 H=g^{1/2}\dzeroh (g^{-1/2}N) =f(x,t;g^{-1/2}\alpha N,
g^{1/2}\alpha^{-1}),
\ee
where $f$ is an arbitrary scalar function of its arguments and
$\alpha$ is an arbitrary scalar density of weight one.  

After using (\ref{Kij}) to replace $\bK$ in (\ref{boxK}) and
designating $g_{ij}$ and $g^{-1/2}\alpha N$ as variables,  the resulting
equation and (\ref{dzeroN}) form a quasi-diagonal hyperbolic system
with principal operators $\dzeroh\Boxh$ and $\dzeroh$.   This system
can also be put in first order symmetric hyperbolic form by
introducing sufficient auxiliary variables and by using the equation
for $R_{00}$ (thus incorporating the Hamiltonian constraint).  
The Cauchy data for the system (in vacuum) are: i) $(\bg,\bK)$ such that the
constraints $R_{0i}=0$, $G^0\mathstrut_0=0$ hold on the initial
slice; ii) $\dzeroh \bK$ such that $R_{ij}=0$ on the
initial slice; and iii) $N>0$ arbitrary on the initial slice.  Note
that the shift $\beta^k(x,t)$ is arbitrary.
Using the Bianchi identities, one can prove 
that this system is fully equivalent to the Einstein equations.  The
constraints of this system propagate in a well-posed causal hyperbolic
way. The restriction on the initial value of $\dzeroh \bK$ 
is what prevents the higher derivative from introducing
spurious unphysical solutions. 

All variables propagate either with characteristic speed zero or the speed 
of light.   The  only variables which propagate at the speed of light
have the dimensions of curvature, and one sees that this is a theory 
of propagating curvature.  

In the above formulation, the shift is arbitrary,
but the lapse is determined by (\ref{dzeroN}). The theory is therefore  
covariant under coordinate transformations which preserve the slicing.
Can the lapse be made arbitrary so that one can simply drop the
generalized harmonic condition (\ref{dzeroN})?  The answer is yes.

By taking another time derivative and adding an appropriate derivative
of $R_{00}$, one finds (in vacuum)
\be
\label{fourth}
\dzeroh \dzeroh R_{ij} - \dzeroh \bar \nabla_i R_{0j} 
-\dzeroh \bar\nabla_j R_{0i}+ \bar\nabla_i \bar\nabla_j R_{00}=
\dzeroh( N\Boxh K_{ij})  +{\cal J}_{ij}=0,
\ee
where ${\cal J}_{ij}$ consists of terms at most third order in
derivatives of $\bg$ and second order in derivatives of $\bK$. 
Together with (\ref{Kij}), these form a system for ($\bg,\bK$) which 
is hyperbolic non-strict in the sense of Leray-Ohya.\cite{LeO}  Both the 
lapse and the shift are arbitrary ($N>0$) and are not dynamical
variables.  The Cauchy data of the previous form (in vacuum) must be 
supplemented for (\ref{Kij}) and (\ref{fourth}) by $\dzeroh \dzeroh \bK$ 
such that $\dzeroh R_{ij}=0$ on
the initial slice.  This guarantees that the system is fully equivalent
to Einstein's theory.  This system does not have a
first order symmetric hyperbolic formulation.

\section*{Acknowledgments} 
A.A., A.A., and J.W.Y. were supported by National Science Foundation
grants PHY-9413207 and PHY 93-18152/ASC 93-18152 (ARPA supplemented).

\end{document}